# Toward a Unified View of Black Hole High Energy States

Michael A. Nowak

JILA, Campus Box 440, Boulder, CO 80309-0440

Electronic Mail: `mnowak@rocinante.colorado.edu`

**ABSTRACT.** We present here a review of high energy (> 1 keV) observations of seven black hole candidates, six of which have estimated masses. In this review we focus on two parameters of interest: the ratio of "non-thermal" to total luminosity as a function of the total luminosity divided by the Eddington luminosity, and the root mean square (rms) variability as a function of the non-thermal to total luminosity ratio. Below $\sim 10\%$ Eddington luminosity, the sources tend to be strictly non-thermal (the so called "off" and "low" states). Above this luminosity the sources become mostly thermal (the "high" state), with the non-thermal component increasing with luminosity (the "very high" and "flare" states). There are important exceptions to this behavior, however, and no steady – as opposed to transient – source has been observed over a wide range of parameter space. In addition, the rms variability is positively correlated with the ratio of non-thermal to total luminosity, although there may be a minimum level of variability associated with "thermal" states. We discuss these results in light of theoretical models and find that currently no single model describes the *full* range of black hole high energy behavior. In fact, the observations are *exactly opposite* from what one expects based upon simple notions of accretion disk instabilities.

## 1. INTRODUCTION

Galactic X-ray sources are typically identified as black hole candidates (BHC) if they have measured mass functions indicating a compact object with $M \gtrsim 3~M_\odot$, or if their high energy spectra ($\sim 1$ keV – 10 MeV) and temporal variability ($\sim 10^{-3} - 10^2$ Hz) are similar to other BHC. A review of the general observations, a number of theoretical models, plus individual descriptions of most of the approximately twenty galactic BHC can be found in Tanaka & Lewin (1995). A review of the timing analyses can be found in van der Klis (1994, 1995) and Miyamoto (1994). These latter reviews attempt to link the energy spectra and timing observations, and they draw analogies to similar observations of neutron stars in low mass X-ray binaries. The purpose of this work is is not to summarize or reanalyze these reviews, but rather to look at a subset of BHC where we have some indication of the compact object's mass and its variability properties. (We make one exception to this list, for reasons which we elaborate below, and include GX 339–4 which has no reliable mass estimate.) We will, however, come to similar conclusions regarding the associated variability and total energetics of each state as are discussed in the reviews of Miyamoto (1994) and van der Klis (1995).

The energy spectra of BHC have been historically labelled based upon observations of the soft X-ray band ($\sim 1-10$ keV). Intense, quasi-thermal flux is referred to as the "high" state. Non-thermal flux in this band, typically a power law with a photon index[1] of $\sim 1.7$, indicates that the BHC is in the "off" state (for extremely low intensity flux) or "low" state (for moderate intensity flux). Note that *in the $\sim 1-10$ keV band* the low state flux is less than the high state flux in this same band.[2] The high state tends to have little variability, with a root mean square (rms) variability of a few percent, whereas the low state tends to have an rms variability of several tens of percent.[3] If a black hole candidate has a quasi-thermal soft X-ray component and significant high energy emission – occasionally modeled as a power law with a photon index $\sim 2.5$ – it is said to be in the "very high" state (*cf.* Miyamoto et al. 1991). The rms variability of the very high state tends to be greater than that of the high state, but less than that of the low state (*cf.* Miyamoto et al. 1992, 1993). Finally, extremely energetic and highly variable non-thermal emission is referred to as the "flare" state, as described below for GS 2023+33. Note that the labels "off", "low", "high", "very high", and "flare" are purely qualitative in nature, but historically have come into popular use – perhaps unfortunately so. No universally agreed upon quantitative definition for these states exist. The distinctions between off, low, and flare, as well as the distinctions between high and very high, are especially murky. Here we will attempt to show somewhat more quantitative, albeit model-dependent, differences between individual BHC observations.

---

[1] Photon index here and throughout shall refer to the photon count rate, such that photon index $\alpha$ implies # photons/keV/s/cm$^2 \propto E^{-\alpha}$, where $E$ is the photon energy.

[2] It is possible, however, for the *bolometric* luminosity to be greater in the low state than in the high state. See the comments by Tanaka & Lewin (1995), and the descriptions of Cygnus X-1 and GX 339–4 below.

Our goal is to arrive at a coherent picture of the off, low, high, very high, and flare states of BHC, and to determine to what extent theoretical models are able to explain this behavior. We will see that although models exist that explain *individual* states, no model fully explains the transition between states. In fact, the transition from a low to high state is opposite from what one expects from simple notions of accretion disk instabilities (*cf.* Piran 1978). An important parameter in many models of these states is $\mathcal{L}_{Tot}$, the ratio of total luminosity to the Eddington luminosity of the compact object (*cf.* Shakura & Sunyaev 1973, Melia & Misra 1993, Luo & Liang 1994, Abramowicz et al. 1995, Chen et al. 1995, etc.). For this reason we choose to study only those BHC where we have an indication of the compact object's mass, and hence are able to define this ratio. In addition, since the various BHC states are partly defined by their variability properties we further restrict ourselves to BHC where detailed timing analyses exist.

We will use the fact that there are two broad classes of theoretical models that are used to fit the spectral components of BHC: the "thermal" and "non-thermal" models. The theoretical basis of the thermal spectra is almost universally accepted to be an optically thick accretion disk (*cf.* Shakura & Sunyaev 1973), although details of individual models differ. The theoretical basis of the non-thermal spectra, by which we mean any spectra that are not simple convolutions of blackbody-like spectra, is much more controversial. We will describe some of the suggested models later, but for now we will take as being "non-thermal" that part of the spectra that is usually fit by a power-law. Thermal and non-thermal components are often fit simultaneously; therefore, we will use this luminosity division as the basis for our discussion. Although this procedure is completely model-dependent, it does provide a quantitative distinction between the BHC states.

For our purposes we will typically compute the non-thermal luminosity from the best fit power-law(s) to the high energy tail(s). Unless additional high energy data indicates otherwise, we shall take the fits to be valid from $1 - 100$ keV (a typical "rollover" for the high energy tails). For the many cases where data exists above 100 keV, we will include this in our luminosity estimates. In addition, we will assume that the non-thermal energy is emitted isotropically. (We will also occasionally use the energetics derived from Comptonization models; *cf.* the description of GX 339–4 below.)

For the thermal component we will use the best-fit "multi-color" or "disk" blackbody (*cf.* Shakura & Sunyaev 1973). Here it is assumed that the emission is from an accretion disk where at each radius, $r$, the disk locally emits as a blackbody with a temperature $T \propto r^{-3/4}$. It is further assumed that the maximum temperature is reached at the inner edge of the accretion disk. The observed flux is then dependent upon the temperature at the inner edge of the disk, $T_{in}$, the radius of the inner edge, $r_{in}$, the inclination of the disk normal with respect to the line of sight, $\theta$, the distance to the source, $d$, and the detector bandpass. Assuming this model is valid for *all* photon energies and not just the energies over which it is fit, the total thermal luminosity of the source is:

$$\mathcal{L}_{Th} \equiv \frac{L_{Th}}{L_{Edd}} = 0.10 \left(\frac{T_{in}}{1 \text{ keV}}\right)^4 \left(\frac{r_{in}(\cos\theta)^{1/2}}{10 \text{ km}}\right)^2 \left(\frac{M_\odot}{M}\right) ,$$

where $M$ is the compact object's mass. Here and throughout, the parameter $\mathcal{L}$ will refer to the luminosity of the source – thermal ($Th$), non-thermal ($NT$), or total ($Tot$) – divided by the Eddington luminosity *for that particular source*. We will generally assume that the disk inclination is given by the average value of $\langle\cos\theta\rangle = 1/2$.

---

[3] The rms variability is defined by the square root of the power spectral density (PSD, *i.e.* the squared Fourier transform) of the photon count rate. The PSD is normalized such that when it is integrated over positive Fourier frequencies, one obtains: $(\langle x^2 \rangle - \langle x \rangle^2)/\langle x \rangle^2$, where $x$ is the photon count rate (*cf.* Miyamoto et al. 1994). The rms is often defined over specific Fourier frequency and energy spectral bandpasses. Whenever possible, we will explicitly state these bandpasses.

As we have mentioned, the above decomposition into thermal and non-thermal components is highly model-dependent, but it gives a rough indication of trends in the sources. The uncertainties in the mass and distance estimates are also large, but as we are only looking for general trends, we will not quote these uncertainties. Suffice to say, they can effect the luminosity estimates by a factor of several.

## 2. DESCRIPTIONS OF SOURCES

We consider seven BHC, three of which are steady X-ray sources (Cyg X-1, LMC X-1, LMC X-3), three of which are transient X-ray sources (GS 1124–683 *a.k.a.* Nova Muscae, GS 2023+33 *a.k.a.* V404 Cyg, GRO J0422+32 *a.k.a.* Nova Persei), and one of which shows some properties of both classes (GX 339–4) (*cf.* Tanaka & Lewin 1995). (Note that we do not include A0620-00, even though it has a well determined mass function, since we have not found any published X-ray power spectra for it.) The assumed masses and distances that we use are as follows:

Cyg X-1 – 6.3 $M_\odot$, 2.5 kpc  (*cf.* Dolan 1992),
LMC X-1 – 6 $M_\odot$, 50 kpc  (*cf.* Hutchings et al. 1987),
LMC X-3 – 9 $M_\odot$, 50 kpc  (*cf.* Cowley et al. 1983),
GS 1124–68 – 6 $M_\odot$, 3 kpc  (*cf.* Remillard et al. 1992, Miyamoto et al. 1993, 1994),
GS 2023+33 – 10 $M_\odot$, 3.5 kpc  (*cf.* Wagner et al. 1992),
GRO J0422+32 – 4.6 $M_\odot$, 2.4 kpc  (*cf.* Kato et al. 1992a,b; Schrader et al. 1993), and
GX 339–4 – 4 $M_\odot$, 4 kpc  (*cf.* the references in Miyamoto et al. 1991).

Note that the mass we use for Cyg X-1 is lower than is often quoted for this object. It is the "best" estimate derived from polarimetric determinations of the inclination of its binary system (Dolan 1992). However, the error bars, as with all the sources, encompass a wide range of masses. The mass of GRO J0422+32 is not determined from a mass function, but rather from a theoretical interpretation of an observed "superhump" period, similar to that seen in cataclysmic variables (*cf.* Kato et al. 1992a,b, Mineshige et al. 1992). The distance to this object is also somewhat uncertain (Schrader et al. 1993). The distances quoted for GS 1124–68 also encompass a wide range. Some references have used 1 or 1.4 kpc instead of 3 kpc (*cf.* Miyamoto et al. 1994). GX 339–4 has no determined mass, and a distance only known to within a factor of two. We choose 4 $M_\odot$ as being consistent with the mass of the other sources, as well as having provided a reasonable fit-parameter for theoretical models of the X-ray variability of GX 339–4 (Nowak 1994). We include this source because it exhibits four distinct spectral states – the off, low, high, and very high states – as well as a wide variety of variability behavior. It covers a wide region of the parameter space in which we are interested, and may provide vital clues linking the behavior of the steady and transient sources. In what follows below we will provide brief descriptions of the data for individual sources.

*Cyg X-1:* During approximately 90% of the observations of Cyg X-1, it has been in a non-thermal state that is reasonably represented by a power law with photon index of $\sim 1.7$ out to $\sim 100$ keV (low state). Above this energy the power law steepens, with Cyg X-1 occasionally showing a flattening of the spectrum between $\sim 1-2$ MeV (*cf.* Ling et al. 1987). Data obtained from Grebenev et al. (1993), Ling et al. (1987), and Ubertini (1991) indicate that $\mathcal{L}_{NT} \sim 0.07$ for the parameters quoted above. In a further proliferation of nomenclature, the non-thermal state has been divided into three sub-states based upon the photon flux in the $45 - 140$ keV band, dubbed in order of increasing photon flux $\gamma_1$, $\gamma_2$, and $\gamma_3$ (*cf.* Ling et al. 1987). Between 10 keV $-$ 10 MeV these states have $\mathcal{L}_{NT} \sim 0.031$, $0.023$, and $0.029$, respectively, with the $\gamma_1$ state being the only one to show flattening in the MeV range. The non-thermal state has also been observed to be highly variable, with an rms $\sim 23\%$ ($0.004 - 70$ Hz, $1.2 - 37$ keV; Miyamoto et al. 1992) and $\sim 44\%$ ($0.015 - 20$ Hz, $4 - 25$ keV; Grebenev et al. 1993). In addition, Cyg X-1 has shown quasi-periodic oscillations (QPO) in its non-thermal state with frequencies of $\sim 0.04$ and $0.07$ Hz and rms $\sim 15\%$ (Kouveliotou et al. 1992a, Ubertini 1994, Vikhlinin et al. 1994a).

Occasionally Cyg X-1 has shown a state with a soft X-ray ($3 - 6$ keV) excess with a power law tail of photon index $\sim 2.2 - 2.7$ extending to high energies (*cf.* Dolan et al. 1979, Ling et al. 1983, Ogarawa et al. 1982). The most recent observations of this "high" state were in 1980. As there are no reliable disk blackbody fits to these data, it is extremely difficult to divide the observations into thermal and non-thermal signatures. If all of the observations between $1 - 12$ keV are ascribed to a disk blackbody emitting at a peak temperature of 1 keV, then $\mathcal{L}_{Th} \sim 0.05$. The data from $12 - 250$ keV, if ascribed completely to non-thermal sources, imply $\mathcal{L}_{NT} \sim 0.014$. Note that these numbers indicate that the bolometric luminosity in the high state is actually slightly lower than the bolometric luminosity in the low state. Unfortunately, there are no reliable observations of the variability of this state. Despite its being touted as a prime example of a source that transits to a high state, Cygnus X-1 shows perhaps the poorest example of what has become known as the high state. Many X-ray transients and GX 339–4 have shown high states which are much more clearly distinct from their non-thermal low states. The high states of these sources have also been observed more frequently and with greater detail than the high state of Cygnus X-1.

*LMC X-1:* Data for this source was taken from Ebisawa et al. (1989) and Hutchings et al. (1987). Energy spectra of the source were well fit by a blackbody with peak temperature of 0.9 keV and a power law tail with photon index 2.3. For the two observations presented in Ebisawa et al. (1989), $\mathcal{L}_{Th} \sim 0.22$, $\mathcal{L}_{NT} \sim 0.025$ for the first observation, and $\mathcal{L}_{Th} \sim 0.26$, $\mathcal{L}_{NT} \sim 0.009$ for the second observation. Measurements gave an rms variability $\sim 11\%$ ($0.004 - 70$ Hz, $1.2 - 15.7$ keV) plus 4% in a 0.08 Hz QPO for the first observation. The second observation had an rms variability $\sim 13\%$ at the same energies and frequencies and showed no QPO. Note that the disappearance of the QPO is consistent with the drop in energy of the hard tail. However, the count rates for these observations were extremely low, and it is unclear whether or not the putative variability detections presented in Ebisawa et al. (1989) were below the effective Poisson noise limit. As presented in Ebisawa et al. (1989), the QPO and continuum variability measurements are inconsistent with one another in the regions that they overlap; therefore, we regard these variability measurements only as upper limits.

*LMC X-3:* The energy spectra for LMC X-3 were similar to the spectra for LMC X-1 (*cf.* Treves et al. 1990, Ebisawa et al. 1993). The spectra were well represented by a disk blackbody with peak temperatures between $0.9 - 1.3$ keV plus a power law with photon indexes between $2.2 - 2.8$. For the observations presented in Treves et al. (1990), the luminosity in these components was $\mathcal{L}_{Th} \sim 0.14$ and $\mathcal{L}_{NT} \sim 0.008$. Treves et al. (1990) claimed that the rms variability was $\sim 6\%$ ($0.004 - 70$ Hz, $1.2 - 37$ keV); however, this variability is entirely consistent with the effective Poisson noise level and should be regarded as an upper limit. The data presented in Ebisawa et al. (1993) showed that the flux of the non-thermal component is apparently uncorrelated with the flux of the thermal component. The thermal component, however, was entirely consistent with a disk accretion model wherein the inner radius, $r_{in}$, remained fixed, but the disk temperature, $T_{in}$, or equivalently the disk accretion rate, varied. No variability measurements, however, are presented in this work.

*GS 1124-683 (a.k.a Nova Muscae):* This source's behavior was characteristic of many X-ray transients.

In the span of less than 10 days, the source flux ($1-6$ keV) rose by more than a factor of 100 (Kitamoto et al. 1992). At its maximum, the X-ray spectrum was well represented by the sum of a disk blackbody with temperature 1 keV and a power law with photon index $\sim 2.5$ (Miyamoto et al. 1993, 1994). The source then decayed with a time constant of $\sim 30$ days for the thermal source and $\sim 13$ days for the non-thermal source. The variability was also seen to decrease as the non-thermal component decreased (Miyamoto et al. 1993, 1994). Approximately 5 months after maximum, the source entered a completely non-thermal phase and continued to decay. Near maximum, a positron annihilation line (with $\mathcal{L} \sim 0.007$) was observed briefly (*cf.* Gilfanov et al. 1993).

GINGA observations (Miyamoto et al. 1993, 1994) of the source indicated that at maximum the source luminosity was given by $\mathcal{L}_{Th} \sim 0.31$ and $\mathcal{L}_{NT} \sim 0.20$, with an rms variability of $\sim 30\%$ ($0.02-60$ Hz, $1.2-37$ keV). Two days later the non-thermal flux dropped to a local minimum ($\mathcal{L}_{Th} \sim 0.25$, $\mathcal{L}_{NT} \sim 0.05$), and the rms variability dropped to 15% (same bandpasses as above). During the first few days after maximum outburst, QPOs between $3-8$ Hz were observed (*cf.* Kitamoto et al. 1992). Approximately two months after maximum outburst, both the non-thermal component and variability were very weak. At one observation the source was observed at $\mathcal{L}_{Th} \sim 0.07$ and $\mathcal{L}_{NT} \sim 0.002$, with rms variability of $\sim 3\%$, while at another observation the source was observed at $\mathcal{L}_{Th} \sim 0.06$ and $\mathcal{L}_{NT} \sim 6 \times 10^{-4}$, with rms variability of $\sim 1\%$ ($1.2-37$ keV, $0.02-60$ Hz; Miyamoto et al. 1993, 1994). When the source first entered into the non-thermal state it had $\mathcal{L}_{NT} \sim 0.06$ with rms variability of $\gtrsim 30\%$ (same bandpasses as above; Miyamoto et al. 1994).

*GS 2023+38 (a.k.a. V404 Cyg):* GS 2023+38 was an unusual transient in that it was one of only two transients that never showed any indication of thermal emission, and in that it showed *extreme* variability on very short time scales (rms $\sim 100\%$). This has been occasionally referred to as the "flare" state. We have obtained data from Sunyaev et al. (1991), Miyamoto et al. (1992), and Tanaka & Lewin (1995). The data indicate that at maximum outburst $\mathcal{L}_{NT} \sim 0.9$. This dropped by a factor of 3 in less than a day. Within 10 days, the luminosity had dropped to $\mathcal{L}_{NT} \sim 0.06$. Within 20 days, the luminosity had dropped to $\mathcal{L}_{NT} \sim 0.03$ and the rms variability was $\sim 20\%$. With such a violent (near Eddington) event with such a rapid decay time perhaps it is reasonable that a quiet, thermal disk structure never appeared to have formed.

*GRO J0422+32 (a.k.a. Nova Persei):* Nova Persei was the second transient to never show any indication of thermal emission; however, it was not as violently variable as V404 Cyg. Data taken from Sunyaev et al. (1992, 1993), Pietsch et al. (1993), and Roques et al. (1994) indicate that at maximum the luminosity never rose beyond $\mathcal{L}_{NT} \sim 0.06 \, (d/2.4 \text{ kpc})^2$. Currently there is not a good estimate for the distance to this object; however, it is unlikely that GRO J0422+32 ever rose above 10% of its Eddington luminosity. (There is, however, some indication of possible excess photon flux around 300 keV; Sunyaev et al. 1993).

As with the other non-thermal sources, there were indications of strong variability. GRANAT found rms variabilities that ranged from $\sim 12\%$ ($0.004-0.04$ Hz, $40-70$ keV), $\sim 9\%$ ($0.004-0.04$ Hz, $70-150$ keV), up to $\sim 50\%$ ($2 \times 10^{-4} - 0.125$ Hz, $150-300$ keV) (*cf.* Sunyaev et al. 1993, Denis et al. 1994). ROSAT observations indicated strong flaring with time scales of $1-10$ s and an rms variability of $\sim 40\%$ ($0.1-2.4$ keV) (Pietsch et al. 1993). In addition, 0.04 and 0.2 Hz QPOs were discovered in the $20-320$ keV bandpass (Kouveliotou et al. 1992b; Sunyaev et al. 1992, 1993; Vikhlinin et al. 1992).

*GX 339–4:* This source shows a wide range of energy spectra and variability, allowing it to be classified as being in either the off, low, high, or very high states. Grebenev et al. (1991) presented data for what could be termed the high and low states of GX 339–4. For these particular observations, the high state was extremely well-represented by a disk blackbody with maximum temperature 0.84 keV, yielding $\mathcal{L}_{Th} \sim 0.03 \, (d/4 \text{ kpc})^2 \, (4 \, M_\odot/M)$. The quoted rms variability was $6\% \pm 3\%$ ($0.04-1$ Hz, $2-60$ keV), although the obtained power spectra were of poor quality. The low state was well-represented by a power-law with photon index 1.7 and luminosity $\mathcal{L}_{NT} \sim 0.045 \, (d/4 \text{ kpc})^2 \, (4 \, M_\odot/M)$. The rms variability was 18% ($0.04-1$ Hz, $2-60$ keV) plus 7% in a 0.8 Hz QPO. Note that for these particular observations, the high state actually had *less* bolometric luminosity than the low state; however, low state observations with much lower luminosity do exist (*cf.* Tanaka & Lewin 1995 for references).

A very high state observation of GX 339–4 was presented in Miyamoto et al. (1991). This state could be reasonably represented by a disk blackbody with maximum temperature of 1 keV and a power-law tail with photon index of 2.5. Miyamoto et al. (1991) also fit the data as a sum of disk blackbody, Comptonized black body, and iron emission line components. This yielded $\mathcal{L}_{Th} \sim 0.075$ $(d/4 \text{ kpc})^2$ $(4 \, M_\odot/M)$ and $\mathcal{L}_{NT} \sim 0.015$ $(d/4 \text{ kpc})^2$ $(4 \, M_\odot/M)$. The rms variability was $\sim 10\%$ $(0.004 - 50 \text{ Hz}, 1.2 - 37 \text{ keV})$ plus 5% in a 6 Hz QPO (similar to the $3-8$ Hz QPO of Nova Muscae). Note that the variability was greater than that of the high state, but less than that of the low state.

## 3. INTERPRETATIONS

The above results are summarized in Figures 1 and 2. Figure 1 presents the ratio of non-thermal to total luminosity plotted vs. the fraction of Eddington luminosity for the observations mentioned in the previous section. This figure also lists some of the labels (low, high, etc.) that have been applied to these observations. Figure 2 plots the fractional rms variability vs. the ratio of non-thermal to total luminosity for the same observations. Note that consistent energy and/or frequency bandpasses *were not* used for the points plotted in Figure 2. However, the points plotted for Nova Muscae, LMC X-1, LMC X-3, and the very high state of GX 339–4 – 10 out of the 17 points shown – all represent roughly similar energy and frequency bandpasses. This is because they all were derived from Ginga observations taken in similar observing modes. In general, Figure 2 shows that variability increases with the ratio of non-thermal to total luminosity.

Figure 1 is rather interesting and somewhat counter to standard theoretical expectations of accretion disk behavior. Below $\mathcal{L}_{Tot} \lesssim 0.07$, the sources tend to be almost entirely non-thermal. Above this luminosity, the sources then become almost entirely thermal. As the luminosity (expressed as a fraction of the Eddington luminosity) is increased even further, the non-thermal component begins to rise again. The so called "very high" state appears to occur at roughly $\mathcal{L}_{Tot} \sim 0.3$, and has substantial non-thermal and thermal components. The one example of a near-Eddington source that we have (GS 2023+33) is once again totally dominated by non-thermal emission. For the case of Nova Muscae, the transition from non-thermal to thermal is consistent with being fairly sharp. For the other sources, it is unclear whether a sharp transition exists, or whether there are merely two broad regions of parameter space in which quasi-equilibrium models can exist. A possible interpretation of this non-thermal to thermal to thermal + non-thermal to non-thermal (with increasing $\mathcal{L}_{Tot}$) behavior is also presented in Figure 1 (solid line with arrows, indicating an evolutionary path for BHC as $\mathcal{L}_{Tot}$ changes that is similar to the one followed by Nova Muscae). If other sources follow such a path, then GRO J0422+32 and GS 2023+33, the only two transients never to have shown a thermal signature, were dominated by non-thermal emission for different reasons. GRO J0422+32 never rose above the transition point at $\mathcal{L}_{Tot} \sim 0.07$, whereas GS 2023+33 rose so far above this break to $\mathcal{L}_{Tot} \sim 1$ that it reached the high luminosity regime where non-thermal emission mechanisms again dominated.

A much more qualitative sketch (which does not show any real data) similar to Figure 1 appears in the review by van der Klis (1995). In his picture of the BHC states, van der Klis shows the very high state occuring at $\mathcal{L}_{Tot} \gtrsim 0.9$, the high state occuring at $0.01 \lesssim \mathcal{L}_{Tot} \lesssim 0.9$, and the low state occuring at $\mathcal{L}_{Tot} \lesssim 0.01$. Though we agree with the general trends, Figure 1 shows that the high state probably occurs over a much narrower range of luminosities ($0.07 \lesssim \mathcal{L}_{Tot} \lesssim 0.3$) and that the very high state occurs at luminosities as low as $\mathcal{L}_{Tot} \sim 0.3$, possibly even as low as $\mathcal{L}_{Tot} \sim 0.1$. This latter fact may be relevant to some radiation-hydrodynamic models of the $\sim 6$ Hz QPO seen in the very high states of Nova Muscae and GX 339-4. These models typically require $\mathcal{L}_{Tot} \sim 0.9$ (*cf.* Miller & Lamb 1992). Only by adopting extreme parameters ($M \sim 2 \, M_\odot$, $d \sim 8$ kpc) can the very high state luminosity of GX 339-4 be made to approach this value.

As we have mentioned, the low state/high state behavior runs counter to simple notions of disk insta-bilities. Simple, optically thick accretion disk models (*cf.* Shakura & Sunyaev 1973) are relatively successful at fitting observations of BHC in their thermal states. However, these models are subject to viscous and thermal instabilities as their luminosity is raised above $\mathcal{L}_{Tot} \sim 0.03 - 0.1$ (*cf.* Lightman 1974, Shakura & Sunyaev 1976, Piran 1978). Some authors have argued that because of these instabilities, disks become optically thin (*cf.* Melia & Misra 1993, Luo & Liang 1994), or develop a hot corona that reflects off of a cool disk (*cf.* Done et al. 1992), or develop hot, optically thick, Comptonizing clouds (*cf.* Sunyaev et al. 1991, 1993, Titarchuk 1994), which then would explain the observed non-thermal spectra. The variability correlated with the increase in the proportion of the non-thermal spectra would then be a consequence of the disk instabilities. Whereas one of these models or some combination of them *may* be the correct explanation for the observed spectra, classical viscous and thermal disk instabilities are unlikely to be the underlying cause. These instabilities are only active where the observations indicate a quiet, thermal source and they are suppressed where the observations indicate a noisy, non-thermal source. To date, no single accretion disk model successfully explains both the non-thermal and thermal spectra, let alone the variability that appears to be correlated with the non-thermal spectra.

We are aware of two suggestions which would allow the non-thermal low states to actually lie at a higher total luminosity than the high states. The first involves the optically thin models of the low state. In these models it is possible to produce spectra wherein a large fraction of the total luminosity is emitted in the MeV range where, until recently, there have been relatively few observations (*cf.* Luo & Liang 1994). This seems unlikely, however, as the observations of Ling et al. (1987) extend to 10 MeV. Including this flux, the total luminosity of Cygnus X-1 is unlikely to be above $\mathcal{L}_{Tot} \sim 0.1$. More recent observations by COMPTEL (McConnell et al. 1994) confirm these limits to the MeV flux, making it very difficult for the optically thin models to "hide" such high luminosities in unobserved regimes.

The second suggestion involves the resurgence in popularity of theoretical models involving advection dominated accretion disks (*cf.* Abramowicz et al. 1995, Chen et al. 1995). In these models the bulk of the accretion energy is advected into the black hole event horizon without radiating. Therefore the total energy (*i.e.* accretion rate) may be higher in the low state while the total *radiated* energy is higher in the high state. The major objection to this scenario, however, is the behavior of the transient BHC. Most of these sources show a clear trend wherein their decay in radiative luminosity is associated with transitions from very high to high, followed by high to low (the one notable exception being V404 Cyg). In the advection dominated scheme the apparent radiative decay would in reality have to be associated with increases in accretion rate. This seems to be unlikely. A large number of X-ray transients *have* shown secondary and tertiary maxima; however, the overall increases in luminosity have been modest. Theoretical models that require relatively small and short lived increases in accretion rate – superimposed on top of a general decline in accretion rate – have been somewhat successful in describing this behavior (*cf.* Chen et al. 1993). Of course it is possible that the accretion phenomena in transient sources is in some way distinct from the accretion phenomena in steady sources.

Uncertainties make comparing X-ray source to X-ray source difficult; however, within a given source we have seen that it is possible for the thermal state occasionally to occur at a slightly lower luminosity than the non-thermal state (*cf.* the descriptions of Cyg X-1 and GX 339–4 above). This raises the question of whether or not there is a hysteresis effect occurring. Does a thermal source tend to stay thermal and a non-thermal source tend to stay non-thermal? An extreme example of this may be GS 2023+33 which after it flared non-thermally remained non-thermal all throughout its rapid decay. Perhaps a more concrete example of this possible hysteresis is the fact that the low state of GX 339–4 is occasionally seen at a slightly higher luminosity than its high state (Grebenev et al. 1991). Unfortunately, Grebenev et al. (1991) does not indicate whether or not these observations occur at the beginnings or endings of low and high state phases. However, this source makes transitions between all of its states on timescales of $200 - 400$ days (*cf.* Harmon et al. 1994), therefore there is the hope of making a more detailed study of its behavior. Recent work with GINGA data (Miyamoto et al. 1995) indicate that these hystereses in GX 339-4 and other sources are indeed truly present, at least in data above 3 keV. We have indicated potential hystereses by the dashed lines in Figure 1 that show alternate evolutionary paths for BHC as $\mathcal{L}_{Tot}$ changes that are similar to the path taken by V404 Cyg and the path possibly taken by GX 339-4. Again, such effects are not accounted for in the disk instability model of low state/high state transitions.

Any theory that hopes to explain these states must not only account for the energy spectra, but also must account for the observed variability. Figure 2 presents the rms variability as a function of the ratio of non-thermal to total luminosity. In general, as the fraction of non-thermal luminosity increases, so does the variability. It is difficult to discern exactly from which components this variability arises; however, recent important work attempts to disentangle the thermal variability from the non-thermal variability in observations of high and very high states (Miyamoto et al. 1994). The results of this study indicate that although the bulk of the short timescale luminosity fluctuations orginate in the non-thermal spectrum, the thermal spectrum fluctuates with an rms $\sim 1-3\%$. In addition, the characterstic PSD of these two components differ. The non-thermal portion of the very high state PSD is flat from $\sim 10^{-2} - 1$ Hz, then rolls over into an $f^{-2}$ power law above this break. That is, the integrated power peaks at 1 Hz timescales (which are – coincidentally? – characteristic accretion disk thermal/viscous timescales). The thermal power seems to be a uniform power-law proportional to $f^{-0.7}$ between $0.01 - 60$ Hz. That is, the integrated power peaks on timescales equivalent to the dynamical timescales of the inner regions of accretion disks. The growth in dominance of the flat-top PSD over the power-law PSD becomes in part a firmer means of distinguishing high states from very high states (*cf.* Miyamoto et al. 1993, 1994). With the imminent launches of the XTE and USA X-ray missions there is hope that we will come to a better understanding of the sources of this variability.

Another likely powerful constraint on models of these states are the observations of black hole QPO. Many observers have attributed these signals – without elaboration – to disk instabilities. No detailed model exists for such a process; however, various purely phenomenological models for both the continuum and QPO power do exist (*cf.* Lochner et al. 1991, Vikhlinin et al. 1994b). It is unclear how disk instabilities would relate to the QPO seen in the non-thermal low state while also relating to the QPO seen in the very high state. If one ignores the 0.08 Hz QPO seen in the high state of LMC X-1 (see the discussion of this source above), then a fairly clean division of QPO frequencies occurs. Low frequency QPO ($< 1$ Hz, typically $\lesssim 0.1$ Hz) are seen in low states, while high frequency QPO ($\sim 6$ Hz) occur in very high states. Anyone who would explain QPO frequencies with disk instabilities needs to address this fact.

## 4. PROSPECTS

The data present both theoretical and observational challenges that must be faced if we are going to arrive at a unified model of black hole high energy states. Of paramount importance is understanding the non-thermal to thermal transition that appears to be occurring at $\mathcal{L}_{Tot} \sim 0.07$. How sharp is this transition? Or are the data more consistent with quasi-equilibrium states existing in broad regions of the $(\mathcal{L}_{Tot}, \mathcal{L}_{NT}/\mathcal{L}_{Tot})$ plane? For the sources that do show spectral transitions, is there really a hysteresis? Answering these questions will involve obtaining better mass and distance determinations for these objects, as well as obtaining as wide an energy coverage as possible. Furthermore, it is important to have multi-wavelength observations of sources as they transit from non-thermal to thermal states *and* from thermal to non-thermal states.

Unfortunately, the only well observed sources that cover large regions of the $(\mathcal{L}_{Tot}, \mathcal{L}_{NT}/\mathcal{L}_{Tot})$ plane are the transients. Observations of the steady sources tend to be grouped in narrow regions of parameter space. LMC X-1 and LMC X-3 are both predominantly thermal sources near the same fraction of Eddington luminosity, whereas Cyg X-1 is predominantly non-thermal for more than 90% of its observations. Is there an intrinsic difference between the emission mechanisms for transient and steady sources? The quasi-steady source GX 339–4 covers the widest range of parameter space. It varies from very weak non-thermal (off state), to strong non-thermal (low state), to thermal (high state), to thermal + non-thermal (very high state). It also shows a wide variety of variability behavior. Unfortunately, its mass is completely unknown, and its distance is not known to within a factor of two. Note that the data for GX 339–4 presented in Figure 1 appears to make a non-thermal to thermal transition at a value of $\mathcal{L}_{Tot}$ lower than that for the other sources. Is this an intrinsic difference of this source, or can this be attributed to uncertainties in its mass and distance? Observations of this source could be important in drawing analogies between the steady and transient sources, as well as being important in exploring the non-thermal to thermal transition.

One could ask whether or not similar non-thermal to thermal behavior (as a function of $\mathcal{L}_{Tot}$) occurs in active galactic nuclei (AGN). One could select a sample of radio-quiet AGN (to insure that the emission is predominantly from a disk, not a jet), with high time-resolution X-ray observations. The *minimum* X-ray variability time scale would be an indication of the *maximum* mass of the central object, and hence provide a *lower limit* to $\mathcal{L}_{Tot}$. The decomposition into thermal and non-thermal luminosities could be performed in a similar manner, except now the thermal component would be predominantly in the optical and UV, rather than the X-ray. We have not attempted to examine such data.

Finally, theorists need to begin to explore models that attempt to unify broad regions of the parameter space. Models are relatively successful at reproducing the energy spectra observations over a narrow range of luminosities (*cf.* the aforementioned references). None successfully bridge the non-thermal to thermal transition. The fact that variability tends to increase with the non-thermal component is very likely an important clue that needs to be considered. In addition, the apparent correlation of low frequency QPO with non-thermal states and high frequency QPO with mixed thermal/non-thermal states also needs to be addressed. Hopefully as both our observations and theories improve, we will arrive at a more coherent picture of these objects.

## ACKNOWLEDGEMENTS

I would like to acknowledge useful conversations with Norm Murray and Chris Thompson, the hospitality of Michel Van der Klis and Brian Vaughan during a brief but productive visit to Amsterdam, and thank Marek Abramowicz and Jean Pierre Lasota for organizing the workshop for which this work was started. This work was partly supported by NASA grant NAGW-4484.

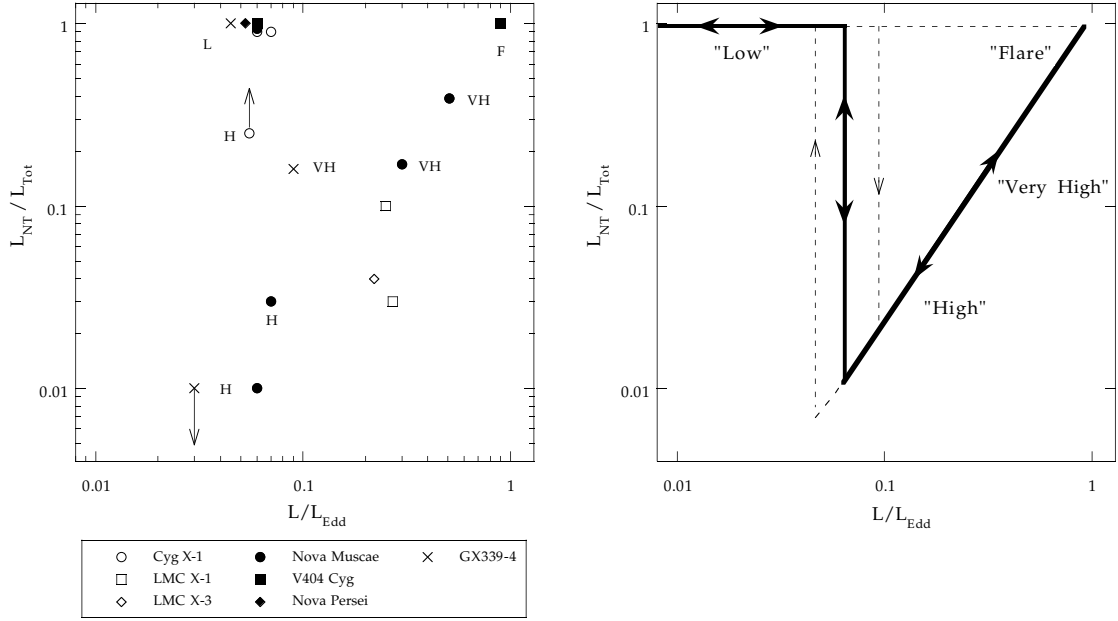

**Figure 1.** *Left–* Ratio of non-thermal to total luminosity vs. fraction of Eddington luminosity for the observations of the 7 BHC described above. Open symbols are for the steady sources, closed symbols are for the transients, and the cross is for GX 339-4. Several of the observations are labelled by their high energy state (L=low, H=high, VH=very high, F=flare). Arrows indicate *highly* uncertain luminosity estimates. *Right–* Hypothetical paths of BHC in the $(\mathcal{L}_{Tot}, \mathcal{L}_{NT}/\mathcal{L}_{Tot})$ plane. Dark lines with arrows indicate the "typical" (?) path of a Nova Muscae-like transient source. The dashed lines indicate possible "alternative" routes taken by other BHC, such as the transient V404 Cyg and the recurrent transient GX 339-4.

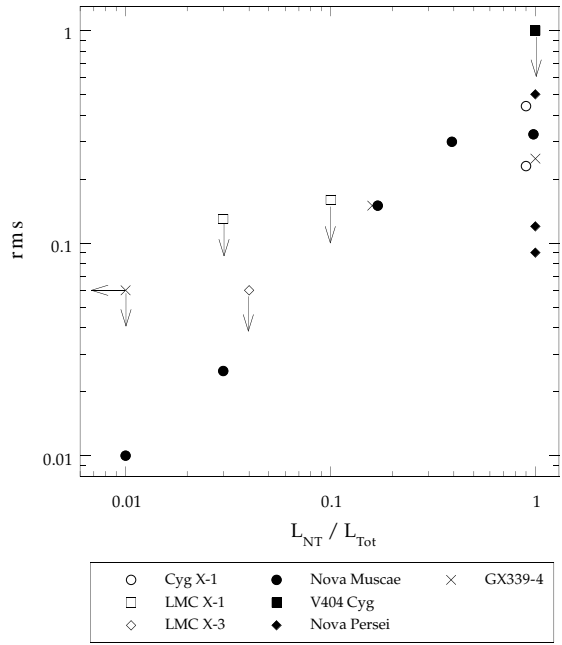

**Figure 2.** Fractional root mean square (rms) variability vs. ratio of non-thermal to total luminosity for the observations of the 7 BHC described above. Open symbols are for the steady sources, closed symbols are for the transients, and the cross is for GX 339-4. Arrows indicate *highly* uncertain luminosity and variability estimates.